\documentstyle[12pt,charm2000]{article}

\def\a{\alpha}

\def\d{\delta} 			\def\D{\Delta}
\def\e{\epsilon}

\def\g{\gamma} 			\def\G{\Gamma}

\def\m{\mu}

\def\p{\pi} 			
\def\q{\theta} 			

\def\s{\sigma} 			\def\S{\Sigma}

\def\ra{\rightarrow}
\def\fr{\frac}
\def\ba{\begin{array}}
\def\ea{\end{array}}
\def\bz{\begin{equation}}
\def\ez{\end{equation}}
\def\by{\begin{eqnarray}}
\def\ey{\end{eqnarray}}

\def\nn{\nonumber}

\newtoks\slashfraction
	\slashfraction={.13}
\def\slash#1{\setbox0\hbox{$\, #1$}
	\setbox0\hbox to \the\slashfraction\wd0{\hss \box0}/\box0}

\begin{document}

\title{Charm Mixing and CP Violation in the Standard Model
\thanks{Presented
at the CHARM2000 Workshop, Fermilab, June 7-9, 1994.}}

\author{Gustavo Burdman \\
{\em Fermi National Accelerator Laboratory, Batavia, Illinois
60510}}

\maketitle
\vspace{-3.6in}
\begin{flushright}
FERMILAB-Conf-94/200 \\
July 1994
\end{flushright}
\vspace{1in}

\setlength{\baselineskip}{2.6ex}

\begin{center}
\parbox{13.0cm}
{\begin{center} ABSTRACT \end{center}
{\small \hspace*{0.3cm}
The Standard Model predictions for $D^0$-$\bar{D}^0$ mixing and CP  
violation
in $D$ decays
are revised. The emphasis is put on obtaining the order of magnitude  
of the
effects. In the case of mixing, the different approaches to the long  
distance
contributions are carefully discussed. The size of CP asymmetries is
discussed in general and some specific calculations are reviewed. The
possibility of using kinematic signals is briefly described.}}
\end{center}

\vskip 0.5truecm

Charm mixing and CP violation are usually thought to be negligibly  
small
in the Standard Model (SM) when compared to the same effects in
the $K$ and $B$ systems. The question of how small is small becomes  
critical
when we consider the possibility of high sensitivity charm  
experiments which
could produce $10^8$ reconstructed $D$ mesons. Although, as we will  
see below,
in most cases the calculations are plagued with strong-interaction
uncertainties making precise predictions impossible, it is of great  
interest
to know at least the order of magnitude of the effects. This allows  
us to
establish the existence or not of windows for the clean observation  
of
new physics beyond the SM. This is particularly true in the case
of mixing.

\section{$D^0$-$\bar{D}^0$ mixing in the Standard Model}
Mixing occurs because the two weak eigenstates $D^0$ and $\bar{D}^0$  
are
not the mass eigenstates. If we neglect CP violation, which as we  
will see
below is a very good approximation  for $D$ mesons, the mass  
eigenstates are
also CP eigenstates and can be written as
\by
|D_1\rangle &=&\fr{1}{\sqrt{2}}\left(|D^0\rangle +|\bar{D}^0\rangle
\right) \nn \\
|D_2\rangle &=&\fr{1}{\sqrt{2}}\left(|D^0\rangle -|\bar{D}^0\rangle
\right). \label{eigen}
\ey
The probability that a $D^0$ meson produced at $t=0$ decays as a  
$\bar{D}^0$
at time $t$ is then given by
\bz
P(D^0\ra\bar{D}^0)=\fr{1}{4}e^{-\G_1t}\left\{1-2e^{-\fr{\D\G}{2}t}\cos 
{\D mt}
+e^{-\D\G t}\right\}, \label{probmix}
\ez
where $\D m=m_2-m_1$ and $\D\G=\G_2-\G_1$ are the mass and lifetime  
differences
in the mass eigenstates. These two quantities determine the ratio of  
``wrong"
final state to ``right" final state in decay modes in which the final  
state
can only be reached by one of the neutral
$D$ meson flavors. This is the case in  semileptonic
decays where we can define
\bz
r_D=\fr{\G (D^0\ra l^-X)}{\G (D^0\ra l^+X)}. \label{defrd}
\ez
This measurable quantity can be expressed in terms of $\D m$ and  
$\D\G$ by
using (\ref{probmix}) and the corresponding expression for the  
unmixed case.
In the limit
\bz
\fr{\D m}{\G}, \fr{\D\G}{\G} \ll 1 \label{limit}
\ez
it takes the simple form
\bz
r_D\approx \fr{1}{2}\left[\left(\fr{\D m}{\G}\right)^2 +
\left(\fr{\D\G}{2\G}\right)^2\right]  \label{rdapprox}
\ez
As we will see,  (\ref{limit}) is a very good approximation.

In the SM $r_D$ is expected to be very small. The question is how
small. In this workshop the possibility of having $10^8$  
reconstructed $D$'s in
various experiments has been
discussed \cite{morr}. It is expected that in some cases a
sensitivity of $10^{-5}$ in $r_D$ could be reached \cite{tliu}.
 Several scenarios for new physics
give contributions to $r_D$ at this level. Therefore it is of great  
interest to
establish at what level the SM contributes. It is not possible to
compute $r_D$ precisely, given the theoretical uncertainties arising  
from long
distance dynamics. Unlike $B^0$-$\bar{B}^0$ mixing, where $r_B$ is  
completely
dominated by the short-distance effects generated by the top  quark,  
the
inherently nonperturbative physics associated with these  
long-distance
effects (e.g. propagation of light quark intermediate states)  is  
potentially
large. In what follows we review the status of our knowledge of the  
short
and long-distance contributions to $\D m$. The lifetime difference  
$\D\G$ is
expected to be of the same order of magnitude as $\D m$ . Given that  
we are
interested in an order of
magnitude estimate we will concentrate on $\D m$.

\subsection{$\D m_D$: Short Distance }
An effective $\D C=2$ interaction is induced, at short distances, by  
one
loop diagrams like the one in Fig.~1, the box diagrams. After the  
loop
integration one obtains \cite{datta}
\bz
{\cal H}_{\mbox{\it eff}}^{\D  
C=2}=\fr{G_F}{\sqrt{2}}\fr{\a}{8\p\sin^2{\q_W}}
|V_{cs}^{*}V_{us}|^2\fr{(m_{s}^{2}-m_{d}^{2})^2}{m_{W}^{2}m_{c}^{2}}
\left({\cal O}+{\cal O}'\right), \label{heff}
\ez
where, in addition to the usual operator
\bz
{\cal O}=\bar{u}\g_\m(1-\g_5)c\bar{u}\g_\m(1-\g_5)c \label{opeusu}
\ez
one has to consider
\bz
{\cal O}'=\bar{u}(1+\g_5)c\bar{u}(1+\g_5)c \label{openew}
\ez
arising from the fact that the mass of the charm quark is not  
negligible.
In (\ref{heff}) we neglect powers of $m_q/m_W$ with $q=d,s$ and the  
$b$ quark
contribution that, although enhanced by a factor of $(m_b/m_W)^2$ is  
largely
suppressed
by the factor $|V_{ub}^{*}V_{cb}|^2$. The GIM mechanism produces the
suppression factor
$(m_{s}^{2}-m_{d}^{2})/m_{W}^{2}$: the effect vanishes in the $SU(3)$  
limit.
The additional suppression $(m_{s}^{2}-m_{d}^{2})/m_{c}^{2}$ comes  
from the
fact that the external momentum, of the order of $m_c$, is  
communicated to the
light quarks in the loop. Both factors explain why the box diagrams
are so small for $D$ mesons relative to the $K$ and $B$ mesons, where  
the GIM
mechanism enters as $m_{c}^{2}/m_{W}^{2}$ and $m_{t}^{2}/m_{W}^{2}$  
and
external momenta can be neglected.

The mass difference generated by the box diagrams is
\bz
\D m=2\langle D^0|{\cal H}_{\mbox{\it eff}}^{\D  
C=2}|\bar{D}^0\rangle,
\label{matem}
\ez
where the matrix elements of the operators ${\cal O}$ and ${\cal O}'$  
can be
parametrized as
\by
\langle D^0|{\cal O}|\bar{D}^0\rangle&=&\fr{8}{3}m_Df_{D}^{2}B_D  
\label{mat1}
\\
\langle D^0|{\cal  
O}'|\bar{D}^0\rangle&=&-\fr{5}{3}\left(\fr{m_D}{m_c}\right)^2
m_Df_{D}^{2}B_{D}'. \label{mat2}
\ey
The vacuum insertion approximation, corresponding to the
saturation of a sum over intermediate states by the vacuum state,  
gives
$B_D=B_{D}'=1$. Corrections to
this simplified approach to the matrix elements are potentially  
large, but are
not expected to change the order of magnitude of the effect.  
Therefore the
box diagram contribution to the mass difference is
\bz
\D m_{D}^{s.d.}\approx 0.5\times 10^{-17}\mbox{\rm GeV}
\left(\fr{m_s}{0.2\mbox{\rm  
GeV}}\right)^4\left(\fr{f_D}{f_\p}\right)^2.
\label{shodis}
\ez
With the $D^0$ lifetime from \cite{pdg}
we have $\G=(1.59\pm0.02)\times 10^{-12}$ GeV. Taking into account  
that the
short-distance contribution to $\D\G$ is of the same order as  
(\ref{shodis}),
we use (\ref{rdapprox}) to obtain the short-distance contribution to  
the
mixing parameter to be
\bz
r_{D}^{s.d.}\approx 10^{-10}-10^{-8},
\ez
which is extremely small.

\subsection{$\D m$: Long Distance}
\subsubsection{Dispersive Approach.}
It has been argued that the fact that the main contributions to  
intermediate
states in $D$ meson mixing come from light quarks  signals the  
presence of
large long-distance effects. They correspond to hadronic intermediate  
states
propagating between the $D$ mesons. It is, in principle, not possible  
to
calculate these effects given their essentially nonperturbative  
character.
However it is crucial to estimate their order of magnitude. In order  
to obtain
it the authors of Ref.~\cite{dght}
make use of dispersive techniques. They consider sets of $n$-particle
intermediate states
related by $SU(3)$. In the $SU(3)$ limit
the contribution from each of these sets must vanish. For instance,  
consider
the intermediate states ivolving
two charged pseudoscalars:  $K^-K^+, \p^-\p^+, K^-\p^+,
K^+\p^-$. Their contribution to mixing comes from diagrams like the  
one in
Fig.~2.
Calculating the loop one typically obtains
\bz
\S(p^2)=A(g)\left[\ln{(-p^2)}+\ldots\right], \label{loop}
\ez
where $p$ is the external momentum and
$A(g)$ depends on the form of the interaction and on the coupling  
$g$.
The ellipses denote constant terms that also depend on the form of  
the vertex.
However the logarithm gives an imaginary part that is related to the  
partial
width of  the on-shell intermediate state. That is, using
\bz
\ln{(-p^2)}=\ln{p^2}+i\p, \nn
\ez
  the relation
\bz
Im\left[\S(p^2)\right]=\G/2 \label{optheor}
\ez
fixes the coefficient of the logarithm. Keeping only this term and  
properly
adding all the charged pseudoscalar states one obtains
\by
\D m_{D}^{l.d.}&\approx &\fr{1}{2\p}\ln{\fr{m_{D}^{2}}{\m^2}}\left[
\G\left(D^0\ra K^-K^+\right)+\G\left(D^0\ra  
\p^-\p^+\right)\right.\nn\\
& &\left.-2\sqrt{\G\left(D^0\ra K^-\p^+\right)\G\left(D^0\ra  
K^+\p^-\right)}
\right], \label{lodis}
\ey
where $\m$ is a typical hadronic scale ($\simeq 1$~GeV). In order to  
get an
estimate for the long-distance effect we would need more information  
on the
doubly Cabibbo-suppressed mode $D^0\ra K^+\p^-$. If we define
\bz
\fr{\G\left(D^0\ra K^+\p^-\right)}{\G\left(D^0\ra K^-\p^+\right)}
=a\times\tan^4{\q_c}, \label{defcasup}
\ez
then in the $SU(3)$ limit one would expect $a=1$. However, a recent  
measurement
by the
CLEO collaboration gives \cite{cleo}
\bz
a=2.95\pm0.95\pm0.95,  \label{a_cleo}
\ez
signaling a possibly large breaking of $SU(3)$. Although the value of  
$\D m_D$
must be proportional to the amount of $SU(3)$ breaking, the value of
(\ref{a_cleo}) does not mean the effect is necessarily large. Large  
$SU(3)$
breaking also occurs in the ratio \cite{pdg}
\bz
\fr{\G\left(D^0\ra K^+K^-\right)}{\G\left(D^0\ra  
\p^+\p^-\right)}\simeq 3,
 \nn
\ez
thus allowing for a partial cancellation of large $SU(3)$ breaking  
effects
in (\ref{lodis}).
In the end the result can be expressed as
\bz
\fr{\D m_{D}^{l.d.}}{\G}\simeq 8\times  
10^{-4}\left(1.4-\sqrt{a}\right)\simeq
-2.5\times 10^{-4}, \label{ldresult}
\ez
where the last number corresponds to taking the central value in
(\ref{a_cleo}). However it can be seen that within the large error  
bars in
(\ref{a_cleo}) the effect is consistent with zero and more data are  
needed.

One could imagine computing, in the same fashion, contributions from  
other
$SU(3)$ related sets of intermediate states: pseudoscalar-vector,
vector-vector,
three pseudoscalars, etc. All of these are proportional
 to the amount of  $SU(3)$
breaking in the set. The relative signs of these contributions are  
unknown and
although
there could be cancellations one would expect the order of magnitude  
to stay
the same.

\subsubsection{Heavy Quark Effective Theory (HQET).}

The applicability of the HQET ideas to $D$-$\bar{D}$ mixing rests on  
the
assumption that the charm quark mass is much larger than the typical  
scale
of the strong interactions. It was first pointed out in  
Ref.~\cite{georgi}
that in this case there are no nonleptonic transitions to leading  
order in
the effective theory since they would require a large momentum  
transferred
from the heavy quark to the light degrees of freedom. This means  
that, in the
effective low energy theory,  mixing is a consequence of matching the  
full
$\D C=2$ theory at the scale $m_c$ with the HQET and then running  
down to
hadronic scales ($\ll m_c$). In other words, there are no new  
operators at
low energy and the only ``long-distance" effects come from the  
renormalization
group running below the matching scale $m_c$. As a consequence, $\D  
m_D$
can be computed in the HQET using quark operators and restricting the
nonperturbative physics only to their matrix elements, which in
Ref.~\cite{georgi} are
estimated using naive dimensional analysis.

First let us consider the four-quark operators generated from the box  
diagrams
by integrating out the $W$'s. These and their matching diagrams in  
the
effective theory are shown in Fig.~3.
 The contribution of these operators
to the mass difference behaves like \cite{georgi}
\bz
\D m_{D}^{(4)}\sim\fr{1}{16\p^2}\fr{m_{s}^{4}}{m_{c}^{2}},  
\label{fq_op}
\ez
where the first factor comes from the loop and $m_d$ is neglected.  
This is
nothing
but the HQET version of the box diagrams.

There will also be higher dimension operators. In principle they will  
be
suppressed by additional powers of $1/m_c$. However, as we see below,  
they
can give important contributions. For instance, six-quark operators  
are
suppressed
by one of such powers. We can think that they arise by ``cutting" one  
of the
light quark lines in the loop in Fig.~4 and then shrinking the  
connecting
line leftover when going to the effective theory given that the  
momentum
flowing through it is large ($\sim m_c$). As a consequence, we get  
rid of
two powers of $m_s$ and the contribution from six-quark operators  
goes like
\bz
\D  
m_{D}^{(6)}\sim\fr{1}{m_c}\fr{m_{s}^{2}}{m_{c}^{2}}\left(m_sf^2\right) 
,
\label{sq_op}
\ez
where the last factor comes from taking the hadronic matrix elements  
and
$f$ is the pseudo-goldstone boson decay constant.

Finally, eight-quark
operators are obtained by cutting the remaining light quark line and
bridging the two four quark pieces with a gluon. The resulting  
contribution
goes like
\bz
\D  
m_{D}^{(8)}\sim\fr{\a_s}{4\p}\fr{1}{m_{c}^{2}}\fr{\left(m_sf^2\right)^ 
2}
{m_{c}^{2}}. \label{eq_op}
\ez
As one can see from (\ref{eq_op}), this is the least GIM-suppressed
contribution. However it is suppressed by $1/m_{c}^{2}$ and most  
importantly
by the  factor $\a_s/4\p$. Relative to the box diagram this is
\bz
\fr{\D m_{D}^{(8)}}{\D m_{D}^{(4)}}\simeq\fr{\a_s}{4\p}\fr{\left(4\p
f\right)^4}
{m_{s}^{2}m_{c}^{2}}\simeq\fr{\a_s}{4\p}\times 20. \label{e_to_f}
\ez
Therefore there is no enhancement due to these operators. In  
Ref.~\cite{georgi}
it is argued that these contributions correspond to the intermediate  
states
taken into account by the dispersive approach. Thus the suppression
factor $\a_s/4\p$
in (\ref{eq_op}) suggests that there are cancellations among the  
different
sets of states.

The six-quark operators give an enhancement of the order of
\bz
\fr{\D m_{D}^{(6)}}{\D m_{D}^{(4)}}\simeq\fr{\left(4\p  
f\right)^2}{m_sm_c}.
\simeq 3\label{s_to_f}
\ez

A complete calculation in this approach,
including $QCD$ corrections to one loop, is performed
in Ref.~\cite{ohl}. Their results can be summarized as
\by
\D m_{D}^{(4)}&\simeq &(0.5-0.9)\times10^{-17}{\rm  
GeV}\left(\fr{m_s}{0.2{\rm
GeV}}
\right)^4 \nn\\
\D m_{D}^{(6)}&\simeq &(0.7-2.0)\times10^{-17}{\rm  
GeV}\left(\fr{m_s}{0.2{\rm
GeV}}
\right)^3 \nn\\
\D m_{D}^{(8)}&\simeq &(0.1-0.6)\times10^{-17}{\rm  
GeV}\left(\fr{m_s}{0.2{\rm
GeV}}
\right)^2. \nn
\ey
In sum, the HQET approach to $\D m_D$ predicts
\bz
\fr{\D m_D}{\G}\simeq (1-2) 10^{-5}. \label{hqet_dmg}
\ez
The uncertainty in (\ref{hqet_dmg}) is mostly due to the uncertainty  
in the
relative signs of the various contributions.
However is clear that HQET predicts no large enhancements with  
respect to the
box diagram, which implies a mixing parameter of the order of
\bz
r_D\approx 10^{-10}-10^{-9}. \label{hqet_rd}
\ez

In conclusion, with the current data on DCSD there seems to be no  
large
disagreement between the dispersive approach of Ref.~\cite{dght} and  
the
HQET estimate of the mixing parameter for $D$ mesons  
\cite{georgi,ohl}.
A conservative upper limit can then be established for the SM  
contribution to
$D^0$-$\bar{D}^0$
mixing to be
\bz
r_{D}^{SM}<10^{-8}. \label{bot_line}
\ez

\section{CP Violation}
In order for CP violation to occur there must be at least two  
amplitudes
interfering with non-zero relative phases. There are two mechanisms  
that
can produce this interference. In the first case the two amplitudes  
correspond
to a $D^0$ decaying as a $D^0$ at time $t$ and a $D^0$ decaying,  
after mixing,
 as a
$\bar{D}^0$ at time $t$, both to the same final state $f$. This is  
called
indirect CP violation and is theoretically clean. That is,
 the hadronic uncertainties
cancel in the asymmetry given that they are the same for both  
amplitudes.
However, as we have seen in the previous section, the mixing  
amplitude is
extremely small in the SM and therefore the induced CP violation
is negligible.

More generally, CP violation can occur directly in the decay  
amplitude. Let us
assume two amplitudes contribute to a given $D$ decay mode. Then
\bz
A_f=A_1e^{i\d_1}+A_2e^{i\d_2}, \label{amp}
\ez
where $A_1$ and $A_2$ are the two amplitudes after factoring out the  
strong
interaction phases $\d_1$ and $\d_2$. When the CP conjugate is taken  
the weak
phases included in $A_{1,2}$ change but the strong phases stay the  
same:
\bz
\bar{A}_{\bar{f}}=A_{1}^{*}e^{i\d_1}+A_{2}^{*}e^{i\d_2}.  
\label{amp_cp}
\ez
The CP asymmetry is then
\bz
a_{CP}=\fr{|A_f|^2-|\bar{A}_{\bar{f}}|^2}{|A_f|^2+|\bar{A}_{\bar{f}}|^ 
2}=
\fr{2Im\left[A_{1}^{*}A_2\right]\sin{(\d_1-\d_2)}}{|A_1|^2+|A_2|^2+
2Re\left[A_{1}^{*}A_2\right]\cos{(\d_1-\d_2)}}. \label{a_cp}
\ez
{}From (\ref{a_cp}) we see that in order to have a nonzero asymmetry  
the
two amplitudes must have different weak as well as strong phases. The
predictions for $a_{CP}$ are then plagued with hadronic uncertainties
coming from the amplitudes and the final-state-interaction phases.

The interesting question is what is the  typical size of the effect  
in the
SM. Before going into the more detailed analysis let us remember that
any CP-violating effect in the SM must be proportional to
the rephasing-invariant quantity
\bz
J=Im\left[V_{ij}V_{kl}V_{ik}^{*}V_{jl}^{*}\right]\label{jarinv}
\ez
for any choice of $i\neq l$ and $j\neq k$. With the current values of
the CKM phases and taking for the CP violating phase $\sin\d=1$ we  
know that
$J\leq 10^{-4}$. From (\ref{a_cp}) we can see that CP asymmetries   
are larger
the
more suppressed is
the mode . For instance, for Cabibbo-suppressed decays we have an  
enhancement
of $\sin^{-2}{(\theta_c)}$ and then an order of magnitude estimate  
for
the asymmetry is
\bz
a_{CP}\sim 10^{-3}. \label{as_ocs}
\ez
In $D$ decays all tree level diagrams contributing to a given final
state have the same CKM matrix element combination. They will  
interfere only
with the one loop diagrams called penguins. Cabibbo-favored $D$ modes  
do not
have penguins and then we are left with Cabibbo-suppressed decays,  
for which
the asymmetry is estimated in (\ref{as_ocs}). However the fact that  
one of the
amplitudes is likely to be much smaller, the penguin in this case,  
largely
reduces the size of the asymmetry. The relative size of the penguin  
to
the tree level diagrams is not a settled issue but one should  
consider
(\ref{as_ocs}) to be on the rather optimistic side unless there is a  
large
enhancement from strong-interaction dynamics, in the same fashion as  
in the
$\D I=1/2$ rule. This possibility is raised in Ref.~\cite{grinstein}.

On the other hand, in $D_s$ decays it is possible to  have two  
tree-level
amplitudes with different weak phases. For instance in $D_{s}\ra K\p$  
the
spectator and annihilation diagrams are proportional to  
$V_{cd}^{*}V_{ud}$ and
$V_{cs}^{*}V_{us}$ respectively. Therefore, if the annihilation  
diagram is
not suppressed relative to the spectator, asymmetries of the order of
(\ref{as_ocs}) are expected.

As was mentioned above, the calculation of the asymmetries involves  
the
knowledge of hadronic matrix elements and strong-interaction phases.
This is done, for instance, in Refs.~\cite{llchau} and  
\cite{buccella}.
In the first case, the relative strong phases are provided by the  
quark
diagrams and final-state interactions are neglected.

 In the work of Ref.~\cite{buccella}, large final-state-interaction  
phases are
provided
by nearby resonances. This tends to give larger asymmetries. The  
typical result
in this case is a few $\times 10^{-3}$. For instance, for the decay
$D^+\ra\bar{K}^*K^+$ $a_{CP}=2.8\times10^{-3}$. In $D_s$ decays the  
most
interesting mode is $K^*\eta '$ with $a_{CP}=-8.1\times10^{-3}$.

In any event, all calculations of direct-CP-violation asymmetries are  
very
uncertain. The  SM can give at most an effect of the order of
$10^{-3}$ but more precise predictions are not possible with our  
current
imprecise knowledge of hadronic physics.

Finally, we  mention the possibility of kinematic
CP-violation signals. For instance, in decays to two vector mesons
$D(p)\ra V_1(k)V_2(q)$ \cite{german,nelson}, it is possible to  
construct CP-odd
correlations of the two polarizations and one of the momenta. A  
triple-product
correlation $\langle k.\e_1\times\e_2 \rangle$ is $T$ odd. However a
non-vanishing value of this quantity is not necessarily a signal of  
CP
violation: the effect could be entirely due to strong-interaction  
phases.
In order to have a truly CP-odd correlation one has to compare with  
the
CP-conjugate state: the sum of
\bz
N_f=\fr{N(k.\e_1\times\e_2 >0)- N(k.\e_1\times\e_2  
<0)}{N_{total}}\label{todd}
\ez
and the corresponding quantity for the CP-conjugate state,  
$N_{\bar{f}}$,
should
vanish if CP is conserved. Similar correlations but for semileptonic  
decays are
discussed in \cite{gergolo}. Another type of kinematic signal can be  
obtained
in neutral three-body decays like $D^0\ra M^+M^-N^0$ \cite{bd}. In  
general the
partial decay rate of a given neutral $D$ flavor need not be  
symmetric in
the energies $E_+$ and $E_-$. However when adding all reconstructed  
neutral
$D$'s from the final state without identifying the $D$ flavor, the  
Dalitz
plot must be symmetric in $E_+, E_-$ unless CP is violated. That is,  
given
the expression
\bz
\G\left[(D^0+\bar{D}^0)\ra M^+M^-N^0\right]=a+b(E_+-E-),  
\label{dp_asym}
\ez
a nonzero value of $b$ signals a net energy asymmetry and therefore  
CP
violation.

In all cases, the kinematic asymmetries are also plagued with  
hadronic
uncertainties as in the case of partial-rate asymmetries
in charged $D$ decays. However it is important that they are taken  
into account
given that in some cases they might be easier to observe.

To summarize, the SM predicts that CP violation in charm decays
proceeds  via the direct mechanism given the small value of
$r_D$. Asymmetries are expected to be at most of order $10^{-3}$ in  
modes
with branching fractions of $10^{-3}$. This implies the need of at  
least
$10^8$ reconstructed $D$'s in order to observe a $3 \s$ effect.

\section{Conclusions}
We have seen that the SM predicts extremely small values for
the mixing parameter $r_D$. The effect, even after including possible
long-distance enhancements, seems to be in the range
$10^{-10}-10^{-8}$.
These effects had been previously overestimated in \cite{wolf} giving
therefore the impression that any observation of $D^0$-$\bar{D}^0$  
mixing
would be contaminated by long-distance dynamics. However this is not
the case. An observation of $D$ mixing at the level of  
$10^{-4}-10^{-5}$,
which is going to be probed at high-sensitivity experiments, would be  
a
signal of new physics \cite{pakvasa}.

On the other hand, CP violation  in the SM might be marginally  
observable
in some cases. Signals from new physics could then be mixed with  
these.
However, there are models where sizeable
asymmetries occur in Cabibbo-favored modes, giving a clear signal  
over the
SM background \cite{pakvasa}.



\begin{thebibliography}{99.}

\bibitem{morr}R.J. Morrison, these proceedings.

\bibitem{tliu}T. Liu, these proceedings.

\bibitem{datta}H. Cheng, Phys. Rev. {\bf D26}, 143 (1982); \\
A. Datta and D. Kumbhakar, Z. Phys. {\bf C27}, 515 (1985).


\bibitem{pdg} Particle Data Group, Phys. Rev. {\bf D45}, 1 (1992).

\bibitem{dght} J.F. Donoghue, E. Golowich, B.R. Holstein and J.  
Trampetic,
Phys. Rev. {\bf D33}, 179 (1986).

\bibitem{cleo}M. Whitherrel, proceeding of the XVI International  
Symposium
on Lepton-Photon Interactions, Cornell University, Ithaca, New York,
August 1993.

\bibitem{georgi}H. Georgi, Phys. Lett. {\bf B297}, 353 (1992).

\bibitem{ohl}T. Ohl, G. Ricciardi and  E.H. Simmons, Nucl. Phys. {\bf  
B403},
605 (1993).

\bibitem{grinstein}M. Golden and B. Grinstein, Phys. Lett. {\bf  
B222}, 501
(1989).

\bibitem{llchau}L.L. Chau and H. Cheng, Phys. Rev. Lett. {\bf 53},  
1037 (1984).

\bibitem{buccella}F. Buccella {\em et al.}, Phys. Lett. {\bf B302},  
319 (1993).

\bibitem{german}G. Valencia, Phys. Rev. {\bf D39}, 3339 (1989).

\bibitem{nelson}J.R. Dell'Aquilla and C.A. Nelson, Phys. Rev. {\bf  
D33}, 80
(1986).

\bibitem{gergolo}E. Golowich and G. Valencia, Phys. Rev. {\bf D40},  
112 (1989).

\bibitem{bd}G. Burdman and J.F. Donoghue, Phys. Rev. {\bf D45}, 187  
(1992).

\bibitem{wolf}L. Wolfenstein, Phys. Lett. {\bf B164}, 170 (1985).

\bibitem{pakvasa}S. Pakvasa, these proceedings.

\end{thebibliography}
\end{document}